\documentclass[12pt,pdftex]{article}

\usepackage[text={6.5in,9.25in}]{geometry}
\usepackage{graphicx}
\usepackage{setspace}
\usepackage{url}
\usepackage{rotating}
\usepackage{color}
\usepackage{amsmath}
\usepackage{amsthm}
\usepackage[utf8]{inputenc}
\usepackage[T1]{fontenc}
\usepackage{graphicx}
\usepackage{amssymb}
\usepackage{epstopdf}
\usepackage{epsfig}
\usepackage{parskip}
\usepackage{graphicx}
\usepackage{subfigure}
\usepackage[authoryear,round]{natbib}
\usepackage{bm}
\usepackage{authblk}

\newcommand{\D}{\mathcal{D}}
\newcommand{\E}{\mathbb{E}}
\newcommand{\e}{\mathrm{e}}
\newcommand{\q}{q}
\newcommand{\ms}{\mathfrak{B}}

\newcommand{\sterling}[1]{\textcolor{black}{#1}}%
\newcommand{\mom}{\vec{\bm{m}}}
\newcommand{\xmin}[1]{X_{\min}^{(#1)}}
\newcommand{\xmax}[1]{X_{\max}^{(#1)}}
\newcommand{\vara}{\alpha}
\newcommand{\varb}{\beta}

\title{Extinction in a branching process:  Why some of the fittest strategies cannot guarantee survival}
\author[1]{Sterling Sawaya\thanks{corresponding author}}
\author[2]{Steffen Klaere}
\affil[1]{\small{Department of Anatomy, University of Otago, Dunedin, \textcolor{blue}{sterlingsawaya@gmail.com}}}
\affil[2]{Department of Statistics and School of Biological Sciences, University of Auckland, \textcolor{blue}{s.klaere@auckland.ac.nz}}

\begin{document}

\vspace{1in}
 \maketitle

%%%%%%%%%%%%%%%%%%%%%%%%%%%%%%%%%%%%%
\begin{abstract}
The fitness of a biological strategy is typically measured by its expected reproductive rate, the first moment of its offspring distribution. However, strategies with high expected rates can also have high probabilities of extinction. A similar situation is found in gambling and investment, where strategies with a high expected payoff can also have a high risk of ruin.  We take inspiration from the gambler's ruin problem to examine how extinction is related to population growth. Using moment theory we demonstrate how higher moments can impact the probability of extinction. We discuss how moments can be used to find bounds on the extinction probability, focusing on $s$-convex ordering of random variables, a method developed in actuarial science.  This approach generates ``best case'' and ``worst case'' scenarios to provide upper and lower bounds on the probability of extinction. Our results demonstrate that even the most fit strategies can have high probabilities of extinction.
\end{abstract}
%%%%%%%%%%%%%%%%%%%%%%%%%%%%%%%%%%%%%
\section{Extinction of a branching process}

Reproduction is necessary for the survival of populations. Populations with high rates of reproduction will often avoid extinction. However, a population may have a high expected reproductive rate but may nevertheless go extinct with near certainty \citep{Lewontin1969}.  For example, populations with large variation in reproductive success can sometimes have a high probability of extinction, even if they have a high expected growth \citep{Orzack1980}.  

Similarly, investors and gamblers can avoid Gambler's Ruin through growth of capital. However, a gambler should not simply apply the strategy with the highest expected growth rate as it may also have a high risk of ruin.  For example, investors can use the Kelly ratio \citep{Kelly} to maximize expected geometric growth of their capital but strict adherence to this ratio can be risky, and playing a more conservative strategy is often recommended \citep{MacLean2010}.

To estimate the probability of Gambler's Ruin, one can use approximations based on moments \citep{Ethier2002, Canjar2007, Hurlimann2005}.  Here we apply these approaches to estimate the probability of extinction in a branching process.  The mathematics of Gambler's Ruin is very similar to that of extinction in a branching process \citep{Courtois2006}.  Both statistical models involve a random variable (payoff/offspring number), resulting in a random walk (change in capital/change in population size), and an absorbing state (ruin/extinction).  Moreover, both processes are assumed to be Markovian, and finding the probability of ruin/extinction involves solving for the root of a convex function.

Here we examine the random variable representing the number of offspring, and investigate how the moments of this random variable are related to the probability of extinction.  We demonstrate an important relationship between these moments and extinction: odd moments favor survival and even moments favor extinction.  The first moment of the offspring distribution, its mean, has the biggest influence on extinction.  However, the first moment alone is not usually informative about extinction probabilities.  In fact, strategies with arbitrarily large first moments can nevertheless go extinct with near certainty.  Some of the ``fittest'' strategies can be highly unlikely to survive.

Using the first few moments of the offspring distribution, one can obtain bounds on the ultimate probability of extinction \citep{Courtois2006, Daley1980}.  These bounds provide ``best case'' and ``worst case'' distributions.  We present these bounds, termed $s$-convex extremal random variables, adapted from actuarial science and research on the gambler's ruin problem \citep{Denuit1997,Hurlimann2005,Courtois2006}.  We find the conditions under which these extremals provide non-trivial bounds.  Using some simple examples, we demonstrate how these methods can be used to compare distributions using their moments.

%%%%%%%%%%%%%%%%%%%%%%%%%%%%%%%%%%%%%
\section{Extinction in the Galton-Watson branching process}

To investigate biological extinction, we use a Galton-Watson branching process in which, at each discrete time interval, every individual generates $i$ discrete offspring with probability $p_{i}$, and zero offspring with $p_{0}$. Without loss of generality we assume that an individual produces its offspring and then dies, so that each individual in a population is restricted to a single generation. The offspring number is a random variable, which we denote by $X$.  Let $n$ be the maximum value of $X$ so that $X$ takes values in the state space $\D_n = \{0,1,2,...,n\}$

At any given time $t$, the size of a population ($Z_t$) is the number of individuals in the branching process. We set $Z_0\equiv1$ unless otherwise specified.  The probability of extinction of a branching process is $\q \equiv \lim_{t \rightarrow \infty} P(Z_{t}=0|Z_{0}=1)$. \sterling{If the starting size of the population is greater than one, then the overall probability of extinction can be defined as}
%------------------------------------
\begin{equation*}
\lim_{t\to\infty}P(Z_t=0|Z_0=N)=\q^N
\end{equation*}
%------------------------------------

\sterling{So we can solve for extinction in the case of $Z_0=1$ and extend the results to larger starting populations if necessary.}

The recursive formula for finding $\q$ can be found through a first step analysis \citep{Kimmel}. The probability that the lineage of a single individual eventually goes extinct is the probability that it dies without offspring $(p_0)$ plus the probability that it produces a single offspring whose lineage dies out $(p_1 \q)$ plus the probability that it produces two offspring whose joint lineages die out $(p_2 \q^2)$, and so on.

This leads to the formal definition of the probability generating function:
%------------------------------------
\begin{equation}\label{ext}
f(q)=\E[q^X]=p_0 + p_1q+p_2q^2+p_3q^3 \dots p_nq^n = \sum_{k=0}^n{p_k\, \q^k}.
\end{equation}
%------------------------------------
The probability of extinction of a branching process starting with a single individual is the smallest root of the equation $f(\q)=\q$ for $\q\in[0,1]$.  The solution $\q=1$ is always a root of \eqref{ext} and is not necessarily the smallest positive root. In some cases, the probability of extinction is trivially obvious.  For instance, if $p_0=0$ individuals always produces at least one offspring, therefore $\q=0$. Furthermore, cases where $\E[X]\le1$ always yield $\q=1$ \citep{Kimmel}.

Inferring the probability of extinction analytically for branching processes with $p_0>0$ and $\E[X]>1$  can be difficult because \eqref{ext} has $n$ complex-valued roots according to the fundamental law of algebra. In the following we illustrate how \eqref{ext} can be seen in terms of moments of the offspring distribution, and discuss how this approach can be used to estimate $\q$.

%%%%%%%%%%%%%%%%%%%%%%%%%%%%%%%%%%%%%
\section{Moments of the branching process}

Let $m_k\equiv\E[X^k]$ denote the $k$th moment of the branching process generator $X$. The first moment, $m_1$, is equivalent to the average offspring number.  Higher moments can be used to obtain other summary statistics of the distribution, such as the variance $\sigma^2=m_2-m_1^2$.

The Laplace transform of \eqref{ext} can be used to (recursively) express extinction in terms of the moments of the branching process
%------------------------------------
\begin{align*}
f(\q)&=\E\left[\q^X\right]=\E\left[\e^{X\,\log \q}\right]\\
\notag&=1+m_1\log \q+m_2\frac{(\log \q)^2}{2}+m_3\frac{(\log \q)^3}{6}+\dots\\
&=\sum_{k=0}^\infty m_k\frac{(\log \q)^k}{k!}
\end{align*}
%------------------------------------
where $m_0=1$. Note that $m_k>0$ for all $k\ge0$. Furthermore, with $\q\in(0,1)$ we have $\log \q<0$. Therefore, even moments increase the probability of extinction while odd moments decrease it. Additionally, if $\q\in(\e^{-1},1)$ then $\log \q \in(-1,0)$ and the series converges with $\log \q$.  Thus, approximations, $f^*(q)$, which take the form
%------------------------------------
\begin{equation*}\label{moment_ext}
f^*(\q)=\sum_{k=0}^{s-1}m_k\frac{(\log \q)^k}{k!}+o\left((\log q)^s\right)
\end{equation*}
%------------------------------------
for $s\ge3$ are only accurate when $\q$ is large and the moments are small.  As $\q \downarrow 0$, the series requires more and more terms to provide accurate approximation.  Therefore, when $q$ is small the first few moments are not necessarily informative about the probability of extinction.

%%%%%%%%%%%%%%%%%%%%%%%%%%%%%%%%%%%%
\section{$s$-Convex orderings of random variables}

Here we demonstrate how the first few moments of the offspring distribution can be used to find bounds on the probability of extinction.  The random variable $X$ is bound by zero and its maximum, $n$, conveniently allowing for $s$-convex ordering \citep{Denuit1997,Hurlimann2005,Courtois2006}. Define the \emph{moment space} for all random variables with state set $\D_n$ and fixed first $s-1$ moments $m_1,\dots,m_{s-1}$ by
%------------------------------------
\begin{equation*}
\ms_{s,n}^{\mom}\equiv \ms (\D_n, m_1, m_2, ...,m_{s-1})
\end{equation*}
%------------------------------------
Since the random variable $X$ is strictly positive, its moment space only contains positive elements.  Further, we are only interested in cases where the mean is greater than 1 so that extinction is not certain.  This provides a moment space with well behaved properties. The study of the moment problem \citep[e.g.,][]{Karlin1957,Prekopa1990} yields an important relationship between consecutive moments on $\ms_{s,n}^{\mom}$ conditional on $m_1\ge1$
%------------------------------------
\begin{equation}\label{moment_rule}
(m_i)^{\frac{i+1}{i}}  \le m_{i+1} \le nm_i
 \end{equation}
%------------------------------------
For two random variables $X$ and $Y$ with state set $\D_n$, we say that if $X$ is smaller that $Y$ in the $s$-convex sense ($X \le_{s-cx}^{\D_n} Y$) then
%------------------------------------
\begin{align*}
\E(X^k)=  \E(Y^k)& \quad\text{for } k=1,2,...,s-1 \\
\E(X^k)\le  \E(Y^k) &\quad \text{for } k \geq s
\end{align*}
%------------------------------------
Minimum and maximum extrema distributions on $\ms_{s,n}^{\mom}$ can be found for any distribution on $\D_n$, with fixed first $s$ moments $m_1,m_2,...,m_s$ \citep{Denuit1997}. The random variables for these distributions are denoted $\xmin{s}$ and $\xmax{s}$ such that
%------------------------------------
\begin{align*}
\xmin{s} \le_{s-cx}^{\D_n} X \le_{s-cx}^{\D_n} \xmax{s} & \quad\text{for all }   X \in \D_n
\end{align*}
%------------------------------------
See \citet{Denuit1997}, \citet{Denuit1999} and \cite{Hurlimann2005} for detailed definitions of s-convexity.  Following the results from these papers, we define the extremal min/max random variables given the first few moments. We begin on $\ms_{2,n}^{\mom}$ with the maximal random variable, $\xmax{2}$, defined as:
%------------------------------------
\begin{equation*}
\xmax{2}=\begin{cases}
0 & \text{with  } p_0=1- \dfrac{m_1}{n} \\
n & \text{with  } p_n=\dfrac{m_1}{n}
\end{cases}
\end{equation*}
%------------------------------------
For $\xmax{2}$ we observe $m_{i+1}=nm_i$, so by \eqref{moment_rule} this can clearly be seen as the maximum extrema.  Intuitively, this is the ``long shot'' distribution on $\D_n$, a worst case scenario.  \sterling{Because the values and respective probabilities of $\xmax{2}$ are known, $q$ can be solved explicitly by finding the least positive root of the generating function:}
%------------------------------------
\begin{equation*}
f(q)=p_0 + p_nq^n
%=1+\log \q+nm_1\frac{(\log \q)^2}{2}+nm_1^2\frac{(\log \q)^3}{6}+\dots\\
\end{equation*}
%------------------------------------
This provides an upper limit on extinction because this generating function will be greater than or equal to the generating function for all other random variables with the same $m_1$ and $n$, on $q \in [0,1]$.

$\ms_{2,n}^{\mom}$ is a very general moment space and the first moment does not often provide much information about an unknown distribution.  Therefore, $\xmax{2}$ is not likely to be a tight upper bound when $n$ is large or unknown.  However, if $m_1$ is near $n$, then the distribution can be fairly well approximated by $\xmax{2}$.  

Unlike $\xmax{2}$, $\xmin{2}$ does not provide a useful bound on the probability of extinction.  $\xmin{2}$ is defined as:
%------------------------------------
\begin{equation}
\xmin{2}=\begin{cases}
\vara &\text{with  } p_{\vara}= \vara + 1 - m_1\\
\vara+1 & \text{with  } p_{\vara+1}= m_1-\vara
\end{cases}\label{x2min}
\end{equation}
%------------------------------------
where $\vara$ is the integer on $\D_n$ such that 
%------------------------------------
\begin{equation*}
\vara <  m_1 \le \vara + 1
\end{equation*}
%------------------------------------
This extremal random variable represents a best case scenario.  However, since $m_1>1$, $\vara$ must be larger than zero and this branching process has no chance of death (i.e. $p_0=0$) and consequently no chance of extinction ($q=0$). Therefore $\xmin{2}$  does not provide a useful bound on the probability of extinction as the bound $q\ge 0$ is obvious.

This bound and all other bounds examined here can be found using discrete Chebyshev systems \citep{Denuit1997}.  However, extremal bounds are perhaps more intuitive for continuous random variables, to which the discrete cases can be seen as similar \citep{Shaked2007, Hurlimann2005, Denuit1999}.  For example, $\xmin{2}$ in the continuous case has only one possible value, $m_1$ with $p_{m_1}=1$. By \eqref{moment_rule} this is clearly an extrema because $(m_i)^{(i+1)/i}  = m_{i+1} = (m_1)^{i+1}$.  In comparison, the discrete case \eqref{x2min} has similar properties.

The following notation helps extending these calculations to higher order systems \citep{Denuit1999b}. Let $w,x,y,z\in\D_n$, and set $m_0=1$. Then:
%------------------------------------
\begin{align*}
m_{j,z}&:=z\cdot m_{j-1}-m_j,\quad j=1,2,\dots;\\
m_{j,z,y}&:=y\cdot m_{j-1,z}-m_{j,z},\quad j=2,3,\dots;\\
m_{j,z,y,x}&:=x\cdot m_{j-1,z,y}-m_{j,z,y},\quad j=3,4,\dots;\\
m_{j,z,y,x,w}&:=w\cdot m_{j-1,z,y,x}-m_{j,z,y,x},\quad j=4,5,\dots
\end{align*}
%------------------------------------
If the first two moments are known, then a tighter upper bound can be found. On $\ms_{3,n}^{\mom}$ the minimal distribution in the $3$-convex sense is given by:
%------------------------------------
\begin{equation*}
\xmin{3}=\begin{cases}
0 &\text{with  } p_{0}=1-p_{\vara}-p_{\vara+1} \\[3mm]
\vara &\text{with  } p_{\vara}= \dfrac{m_{2,\vara+1}}{\vara}\\[3mm]
\vara +1 &\text{with  } p_{\vara+1}=\dfrac{-m_{2,\vara}}{\vara+1}
\end{cases}
\end{equation*}
%------------------------------------
where 
%------------------------------------
\begin{equation*}
\vara<\frac{m_2}{m_1}\le\vara+1.
\end{equation*}
%------------------------------------
This bound is already known in the branching process literature \citep{Daley1980}.  Similar to $\xmax{2}$, the extremal random variable $\xmin{3}$ represents a worst case scenario, this time using two moments.  The root of the equation
%------------------------------------
\begin{equation}\label{pgf_x4}
f(q)=q=p_0+p_{\vara}q^{\vara}+p_{\vara+1}q^{\vara+1}
\end{equation}
%------------------------------------
provides an upper bound to the probability of extinction, so that \eqref{pgf_x4} has greater values at any $q\in[0,1)$ than the probability generating functions of any other random variable in $\ms_{3,n}^{\mom}$.

In contrast to $\xmax{2}$, the minimum extrema on $\ms_{3,n}^{\mom}$ yields the upper limit for the probability of extinction. The alternation between minimum and maximum for the worst case scenarios is due to the convexity of \eqref{ext}. Again, this extrema is perhaps more intuitive in the continuous sense, in which 
%------------------------------------
\begin{equation*}
X_{\min,\  \text{cont.}}^{(3)}=\begin{cases}
0 &\text{with  } p_{0}=1-p_{m_2/m_1} \\
\dfrac{m_2}{m_1} &\text{with  } p_{m_2/m_1}= \dfrac{(m_1)^2}{m_2}
\end{cases}
\end{equation*}
%------------------------------------
In this case, successive moments simply grow by $m_2/m_1$, so that $m_{i+1}=m_i (m_2/m_1)$, providing a clear minimum on $\ms_{3,n}^{\mom}$.  And, as was the case for the minimum on $\ms_{2,n}^{\mom}$, the discrete minimum extrema on $\ms_{3,n}^{\mom}$ has similar properties to the continuous minimum extrema.

For both $\ms_{2,n}^{\mom}$ and $\ms_{3,n}^{\mom}$ the discrete cases are simply discretization of the continuous case. However, this is not necessarily the case for higher moment spaces \citep{Courtois2006}. While the continuous cases provide more intuitive extrema, derivation of the discrete case for higher moments is not as simple as deriving the continuous case and discretizing.

Next, we examine the maximum extrema on $\ms_{3,n}^{\mom}$:
%------------------------------------
\begin{equation*}
\xmax{3}=\begin{cases}
\vara &\text{with  } p_{\vara}=  \dfrac{m_{2,n,\vara+1}}{n-\vara}\\[3mm]
\vara+1 &\text{with  } p_{\vara+1}=  \dfrac{-m_{2,n,\vara}}{n-\vara-1}\\[3mm]
n &\text{with  } p_{n}=1-p_{\vara}-p_{\vara+1}
\end{cases}
\end{equation*}
%------------------------------------
where
%------------------------------------
\begin{equation*}
\vara <  \frac{nm_1-m_2} {n-m_1} \le \vara +1
\end{equation*}
%------------------------------------
Since $\xmax{3}$ can only provide non-trivial information about $q$ if $p_0>0$, this extremal distribution is only informative about extinction when $\vara=0$ and $p_{\vara}>0$, which is the case whenever $nm_1-m_2<n-m_1$. Although this requirement may appear restrictive, some classes of distributions have simple rules under which  $\xmax{3}$ is informative.  For example, for binomial distributions, $B_{n,p}$, $\xmax{3}$ will provide a non-zero lower bound if $1/n<p\le 1/(n-1)$.

We move on to $\ms_{4,n}^{\mom}$. The use of three moments can improve bounds on the probability of extinction, but as with all of the maximal random variables, $\xmax{4}$ requires the knowledge of the  maximum, $n$.  $\xmax{4}$ is defined as:
%------------------------------------
\begin{equation*}
\xmax{4}=\begin{cases}
0 &\text{with  } p_{0} = 1-p_{\vara}-p_{\vara+1}-p_n\\[3mm]
\vara &\text{with  } p_{\vara}= \dfrac{m_{3,n,\vara+1}}{\vara(n-\vara)} \\[3mm]
\vara+1&\text{with  } p_{\vara+1}=\dfrac{-m_{3,n,\vara}}{(\vara+1)(n-\vara-1)} \\[3mm]
n &\text{with  } p_{n}=\dfrac{m_{3,\vara,\vara+1}}{n(n-\vara)(n-\vara-1)}
\end{cases}
\end{equation*}
%------------------------------------
where
%------------------------------------
\begin{equation*}
\vara <  \frac{m_2n-m_3} {m_1n-m_2} \le \vara +1
\end{equation*}
%------------------------------------
While this is a potential improvement to the lower bound given by $\xmin{3}$, the improvement is sometimes negligible.  As $n \to \infty$, the  difference between $\xmax{4}$ and $\xmin{3}$ vanishes because 
%------------------------------------
\begin{align*}
\lim_{n \to \infty} &\ \frac{m_2n-m_3} {m_1n-m_2} = \frac{m_2} {m_1}
\end{align*}
%------------------------------------
and because with $p_n \to 0$, the generating function for  $\xmax{4}$ is identical to \eqref{pgf_x4}.  So, like the first moment, the third moment is uninformative about extinction when $n$ is unknown, unless assumptions are made about the distribution \citep[see, e.g.,][]{Daley1980, Ethier2002}.

The minimal extrema for $\ms_{4,n}^{\mom},\,\xmin{4}$ is given by
%------------------------------------
\begin{equation*}
\xmin{4}=
\begin{cases}
\vara,&\text{with }p_\vara=\dfrac{m_{3,\varb,\varb+1,\vara+1}}{(\varb-\vara)(\varb+1-\vara)}\\[3mm]
\vara+1,&\text{with }p_{\vara+1}=\dfrac{-m_{3,\varb,\varb+1,\vara}}{(\varb-\vara)(\varb-1-\vara)}\\[3mm]
\varb,&\text{with }p_\varb=\dfrac{m_{3,\vara,\vara+1,\varb+1}}{(\varb-\vara)(\varb-1-\vara)}\\[3mm]
\varb+1,&\text{with }p_{\varb+1}=\dfrac{-m_{3,\vara,\vara+1,\varb}}{(\varb-\vara)(\varb+1-\vara)}
\end{cases}
\end{equation*}
%------------------------------------
where $\vara$ and $\varb$ are given by
%------------------------------------
\begin{equation*}
\vara<\dfrac{m_{3,\varb,\varb+1}}{m_{2,\varb,\varb+1}}\le\vara+1,\quad
\varb<\dfrac{m_{3,\vara,\vara+1}}{m_{2,\vara,\vara+1}}\le\varb+1.
\end{equation*}
%------------------------------------
Again, this bound is only useful if $p_0>0$.  Unfortunately there is no short form equation to identify  which spaces $\ms_{4,n}^{\mom}$ fit this requirement.  However, one can easily determine if a given $\ms_{4,n}^{\mom}$ has a useful $\xmin{4}$.  Assuming $\vara=0$, $\widehat{\varb}$ is simply bound by
%------------------------------------
\begin{equation*}
\widehat{\varb}<\frac{m_3-m_2}{m_2-m_1}\le \widehat{\varb}+1
\end{equation*}
%------------------------------------
And if $m_{3,\widehat{\varb},\widehat{\varb}+1}< m_{2,\widehat{\varb},\widehat{\varb}+1}$, then the bound is useful because the resulting $\xmin{4}$ has $p_0>0$.  Alternatively, if $m_{3,\widehat{\varb},\widehat{\varb}+1} \ge  m_{2,\widehat{\varb},\widehat{\varb}+1}$ the the supports for $\xmin{4}$ have $p_0=0$ and consequently $q=0$.

If the first four moments are known, the extremal variable $\xmin{5}$ can be obtained.  Its distribution takes a simple form, but the equations used to find its values and relative probabilities are relatively large.  From \citet{Hurlimann2005}, $\xmin{5}$ is defined as:
%------------------------------------
\begin{equation*}
\xmin{5}=\begin{cases}
0&\text{with  } p_{0}=1-p_\vara-p_{\vara+1}-p_\varb-p_{\varb+1} \\[3mm]
\vara &\text{with  } p_{\vara}=\dfrac{m_{4,\varb,\varb+1,\vara+1}}{\vara(\varb-\vara)(\varb+1-\vara)} \\[3mm]
\vara+1 &\text{with  } p_{\vara+1}=\dfrac{-m_{4,\varb,\varb+1,\vara}}{(\vara+1)(\varb-\vara)(\varb-1-\vara)} \\[3mm]
\varb &\text{with  } p_{\varb}=\dfrac{m_{4,\vara,\vara+1,\varb+1}}{\varb(\varb-\vara)(\varb-1-\vara)} \\[3mm]
\varb+1 &\text{with  } p_{\varb+1}=\dfrac{-m_{4,\vara,\vara+1,\varb}}{(\varb+1)(\varb-\vara)(\varb+1-\vara)}
\end{cases}
\end{equation*}
%------------------------------------
where
%------------------------------------
\begin{equation}\label{eq.ineq}
\vara<\frac{m_{4,\varb,\varb+1}}{m_{3,\varb,\varb+1}}\le\vara+1,\quad
\varb<\frac{m_{4,\vara,\vara+1}}{m_{3,\vara,\vara+1}}\le\varb+1.
\end{equation}
%------------------------------------
\citet{Courtois2006} proposed that there is no analytic form to directly obtain $\vara$ and $\varb$ for $\xmin{5}$. They showed this by disproving the intuitive idea that the discrete support encloses the continuous support.  \sterling{Thus, to find $\vara$ and $\varb$, iteratively search $D_n$ until both inequalities are satisfied.}

%\steffen{Thus it is more useful to initialise with $\vara=0$, and iteratively using \eqref{eq.ineq} until both values satisfy the inequalities.}

\citet{Hurlimann2005} also presents a form for the upper extremal variable  in $\ms_{5,n}^{\mom}$. The process $\xmax{5}$ is defined as:
%------------------------------------
\begin{equation*}
\xmax{5}=
\begin{cases}
\vara,&\text{with }p_\vara=\dfrac{m_{4,n,\varb,\varb+1,\vara+1}}{(\varb-\vara)(\varb+1-\vara)(n-\vara)}\\[3mm]
\vara+1,&\text{with }p_{\vara+1}=\dfrac{-m_{4,n,\varb,\varb+1,\vara}}{(\varb-\vara)(\varb-\vara-1)(n-\vara-1)}\\[3mm]
\varb,&\text{with }p_\varb=\dfrac{m_{4,n,\vara,\vara+1,\varb+1}}{(\varb-\vara)(\varb-\vara-1)(n-\varb)}\\[3mm]
\varb+1,&\text{with }p_{\varb+1}=\dfrac{-m_{4,n,\vara,\vara+1,\varb}}{(\varb-\vara)(\varb+1-\vara)(n-\varb-1)}\\[3mm]
n,&\text{with }p_n=1-p_\vara-p_{\vara+1}-p_\varb-p_{\varb+1}
\end{cases}
\end{equation*}
%------------------------------------
where
%------------------------------------
\begin{align*}
\vara<\frac{m_{4,n,\varb,\varb+1}}{m_{3,n,\varb,\varb+1}}\le\vara+1,\quad
\varb<\frac{m_{4,n,\vara,\vara+1}}{m_{3,n,\vara,\vara+1}}\le\varb+1.
\end{align*}
%------------------------------------
As was the case for $\xmin{4}$, one can determine if $\xmax{5}$ has $p_0>0$ by assuming $\vara=0$ and solving for $\widehat{\varb}$ with
%------------------------------------
\begin{equation*}
\widehat{\varb} < \frac{m_{4,n,0,1}}{m_{3,n,0,1}} \le \widehat{\varb}+1
\end{equation*}
%------------------------------------
If the resulting $\widehat{\varb}$ in the inequality $m_{4,n,\widehat{\varb},\widehat{\varb}+1}<m_{4,n,\widehat{\varb},\widehat{\varb}+1}$ holds,  the bound for $\xmax{5}$ is informative.

\sterling{All $\xmax{j}$ extrema rely on the maximum offspring number, $n$.  Similar to $\xmax{4}$, when $n$ is unknown or infinity $\xmax{5}$ goes to the minimum on the lower moment space, here $\xmin{4}$.  Thus if $n$ is unknown, $\xmax{j}$ goes to $\xmin{j-1}$, at least for the cases examined here. }

\sterling{The Chebychev approach can be used to extend this approach to higher moments \citep{Hurlimann2005}, however we do not believe this would be worthwhile for two reasons.  First, moments above the fourth are rarely used, and higher moments can be difficult to estimate.  Further, the equations for the supports and probabilities for moments above the fourth become immense, and calculating their values for a given set of moments may be challenging.}

%\steffen{It is worth noting that all $\xmax{j}$ variables need the knowledge of the maximal support value, $n$. Further, as $n\to\infty$ $\xmax{j}$ goes to $\xmin{j-1}$. Thus, if $n$ is unknown, e.g., if one only has a sample from the population, one could use $\xmin{j},\,j=2,\dots,5$ to find bounds on the extinction probabilities for the sample.}

%\steffen{While the Chebychev approach can be extended to more than four moments, it is unlikely that the increase in computational effort makes this task appealing. Further, in real-world applications, the estimates for the moments become more unreliable with increasing $k$, and hence the more moments one estimates the less reliable the estimate of the extremal distribution gets.}

%%%%%%%%%%%%%%%%%%%%%%%%%%%%%%%%%%%%%
\newpage
\section{Examples}

Here we discuss some example distributions, graph their generating functions, and also graph generating functions for the extremal distributions.  The plot of the probability generating function, $f(q)$, on $q \in (0,1)$ is a useful way to visualize how the moments are related to extinction.  The probability generating function takes the value $p_0$ at $q=0$.  At small $q$, $f(q)$ has a slope of approximately $p_1$.  In this part of the function, when $\q$ is small, there can be a weak relationship between $f(q)$ and moments.  In comparison, when $q$ is close to 1, the moments are closely related to $f(q)$.  For example $f'(1)=m_1$.  Higher moments begin to influence the function as $q$ moves away from 1. 

The probability of extinction of a process is found when $f(q)=q$, i.e at the intersect between its probability generating function $f(q)$ and the diagonal $q$. Thus, processes with a high probability of extinction will cross the diagonal near $q=1$, in the domain of $q$ in which the probability generating function is often closely related to its first few moments.

%===================================
\begin{figure}[!ht]
\subfigure[Binomial distribution $\text{Bin}_{20,0.1}$]{
	\label{fig.bindist}
	\begin{minipage}{.5\textwidth}
	\includegraphics[width=\textwidth]{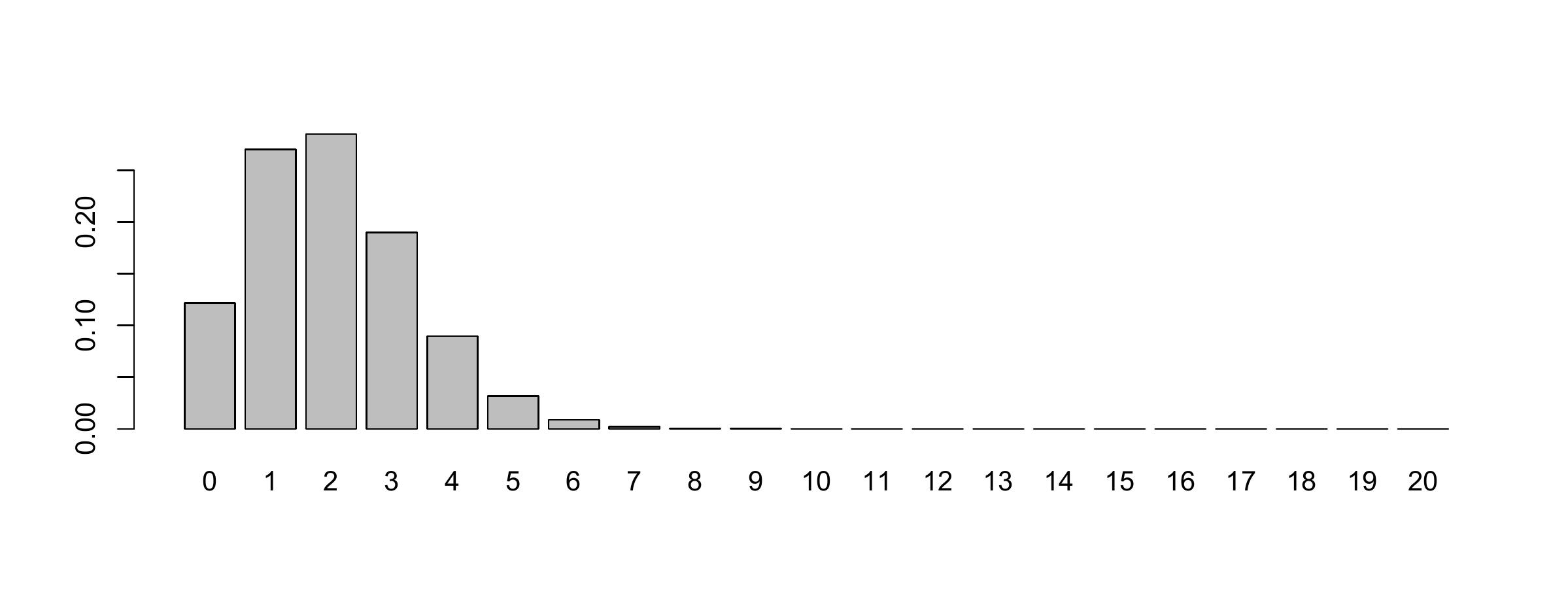}\\
	\includegraphics[width=\textwidth]{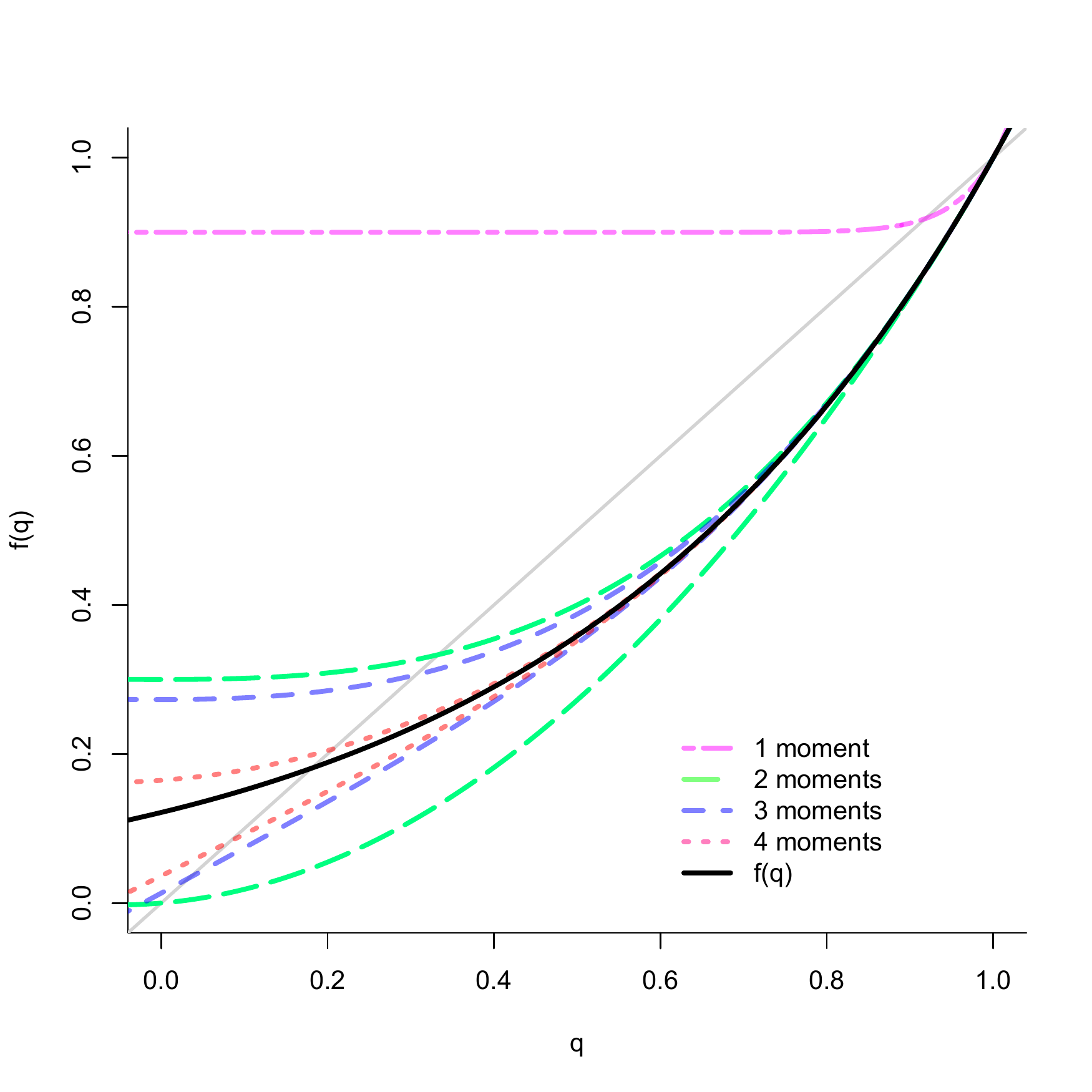}
	\end{minipage}
}
\subfigure[Truncated geometric distribution]{
	\label{fig.geodist}
	\begin{minipage}{.5\textwidth}
	\includegraphics[width=\textwidth]{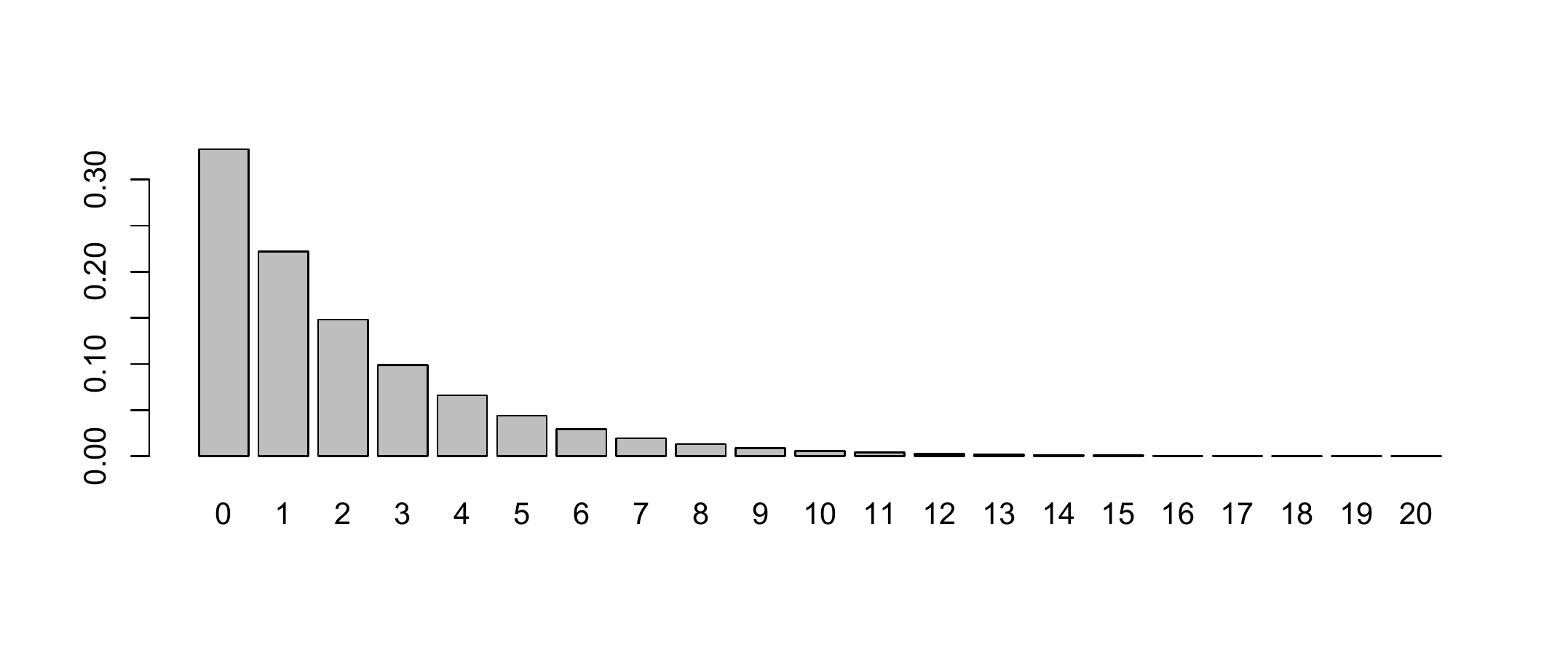}\\
	\includegraphics[width=\textwidth]{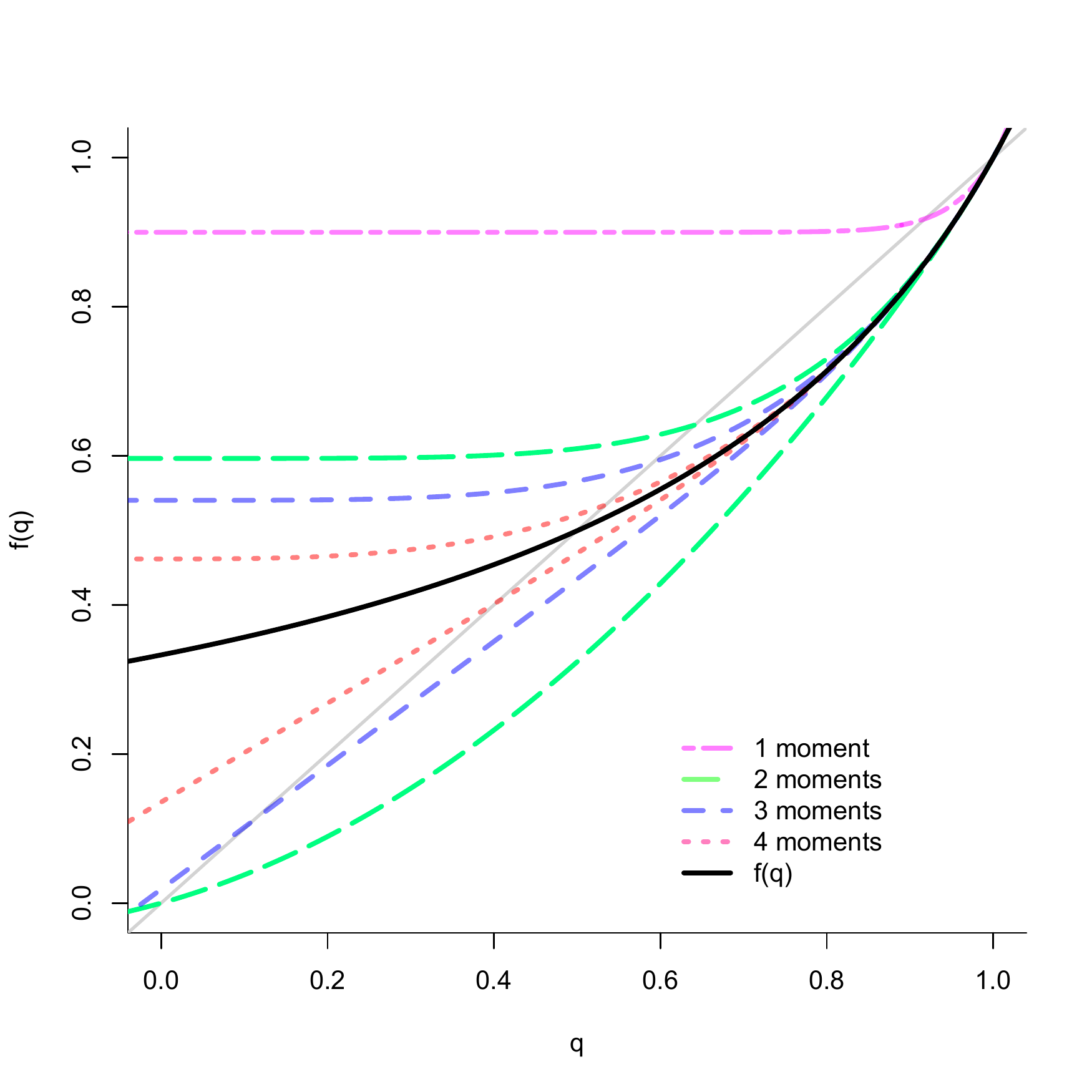}
	\end{minipage}
}
\caption{\label{fig.pgf}Probability generating functions, $f(q)$, for a Binomial and a truncated Geometric distribution. Both distributions have a maximal value of $20$, and a mean of $2$. The offspring distribution is shown above each  graph of the respective probability generating functions.  The probability generating functions for the extremal distributions are also plotted, and can be found above and below the plot of the generating functions.}
\end{figure}
%===================================

Plotting the probability generating functions for the extremal distributions helps demonstrate why they act as bounds on extinction.  In these examples (Figure \ref{fig.pgf}), we compare two distributions with identical first moment and maximum ($m_1=2$, $n=20$), i.e. both distributions are in $\ms_{2,20}^2$. In particular, we look at a binomial distribution, Figure \ref{fig.bindist}, and a truncated geometric distribution, Figure \ref{fig.geodist}. For each of these plots we also plot the generating functions for some of the extremal distributions.  The extremal distributions provide clear bounds: best case extrema are found below the plot of the generating function, worst case extrema are found above.  For example, the extremal distribution based on one moment, $\xmax{2}$, provides an upper bound on the probability of extinction, and can be seen as the upper line in both plots.  Because they share an identical first moment and maximum, $\xmax{2}$ is the same for both distributions.  Clearly, one moment does not provide a good bound in these examples. As more moments are used, the bounds become tighter.  The extrema using four moments provide relatively accurate upper and lower bounds for both examples.

The lower bounds provide the best case extrema, which are useful in both cases only when three or four moments are known.  The lower bound using two moments is not useful here in either case, as its probability generating function crosses the diagonal at zero so its probability of extinction is zero.  The lower bound using only one moment was not included because its generating function is trivial and always uninformative about extinction.

%\steffen{Tables \ref{ext.bin} and \ref{ext.geo} show the extinction probabilities and supports for both example distributions and their respective bounding processes, indexed by the number of moments they share.}
%reference to tables added to the paragraph
Importantly, these examples demonstrate why higher moments are often necessary to compare strategies. These two distributions have identical first moments ($m_1=2$) so classically their fitness value would be equal. However, the binomial example is more likely to survive. If entire distributions are known, then extinction probabilities can be calculated explicitly. If instead enough moments are known, one can nevertheless conclude that the truncated geometric example is inferior.  Compare the extremal distributions when four moments are known, paying attention to where they cross the diagonal.  \sterling{The value at the intersect is the probability of extinction for the extrema, which we display in Table 1 and Table 2, respectively for the binomial example and the truncated geometric example. Using four moments, the best case for the truncated geometric example (0.404, Table 2) is worse than the worst case for binomial example (0.207, Table 1). In fact, the worst case for the binomial example using two moments (0.333) is already better than the best case for the truncated geometric using four moments (0.404). These examples highlight how moment spaces can be used to compare branching processes.}

%===================================
\begin{table}[h]
\begin{center}
\begin{tabular}{|r||r|r|r|r|}
\hline
&    1 moment &    2 moments &    3 moments & 4 moments \\\hline\hline
supports   & $\{2\}$&$\{1,2,20\}$&$\{0,1,3,4\}$&$\{0,1,3,4,20\}$\\\hline
lower bound (best case) &$0.000$&$0.000$&$0.034$&$0.083$\\\hline\hline
supports   & $\{0,20\}$&$\{0,2,3\}$&$\{0,2,3,20\}$&$\{0,1,2,4,5\}$\\\hline
upper bound (worst case) &$0.918$&$0.333$&$0.306$&$0.207$\\\hline
\end{tabular}
\caption{\label{ext.bin}Extinction probabilities and supports for the extremal distributions of the Binomial example $B_{20,0.1}$.  The actual probability of extinction for this process is 0.181.}
\end{center}
\end{table}
%===================================

%===================================
\begin{table}[h]
\begin{center}
\begin{tabular}{|r||r|r|r|r|}
\hline
&    1 moment &    2 moments &    3 moments & 4 moments \\\hline\hline
supports   &  $\{2\}$&$\{1,2,20\}$&$\{0,1,7,8\}$&$\{0,1,6,7,20\}$\\\hline
lower bound (best case) &$0.000$&$0.000$&$0.110$&$0.404$\\\hline\hline
supports   &  $\{0,20\}$&$\{0,4,5\}$&$\{0,4,5,20\}$&$\{0,3,4,11,12\}$\\\hline
upper bound (worst case) &$0.918$&$0.641$&$0.592$&$0.534$\\\hline
\end{tabular}
\caption{\label{ext.geo}Extinction probabilities and supports for the extremal distributions of the truncated Geometric example.  The actual probability of extinction for this process is 0.499.}
\end{center}
\end{table}
%===================================

%\steffen{In another example, assume a project wants to estimate the probability of extinction in a population by sampling the offspring number of 25 females in the community. We simulate potential results using a Poisson distribution with parameter $\lambda=2.5$:
%------------------------------------
%\begin{equation*}
%\{2,3,3,1,6,3,1,2,2,4,2,5,0,1,0,2,0,1,0,5,4,2,1,3,6\}.
%\end{equation*}
%------------------------------------
%We estimate the first four moments from the sample, and compute the extremal distributions $\xmin{j},\,j=2,3,4,5$ for these values, summarized in Table \ref{ext.sample}. We find that the sample indicates a probability of extinction around $0.216$ with its extremal distribution suggesting a range between $0.04$ and $0.237$. An estimation of the ``true'' value for the Poisson distribution $\pi_{2.5}$ returns an extinction probability of around $0.107$. Thus, we can use the extremal processes to generate an uncertainty interval for the extinction probability of the true value.  
%===================================
%\begin{table}[h]
%\begin{center}
%\begin{tabular}{|r||r|r|r|r|r|}
%\hline
%Moments&    sample&    $2$&    $3$&    $4$&$5$\\\hline\hline
%supp   & $\{0,1,\dots,6\}$&  $\{2,3\}$&$\{0,3,4\}$&$\{0,1,4,5\}$&$\{0,2,3,5,6\}$\\\hline
%prob  &$0.216$&$0.000$&$0.373$&$0.040$&$0.237$\\\hline
%\end{tabular}
%\caption{\label{ext.sample}\steffen{Extinction probabilities and supports for the extremal processes for the above sample.}}
%\end{center}
%\end{table}
%===================================
%}

%%%%%%%%%%%%%%%%%%%%%%%%%%%%%%%%%%%%%
\newpage
\section{Discussion}

The work here is intended to highlight the relationship between the moments of the offspring distribution and the probability of extinction.  Extinction can be defined in terms of moments, but the first few moments are only informative about extinction under certain conditions.  But, no matter these conditions there exists an interesting relationship with even and odd moments:  high even moments favor extinction, high odd moments favor survival.  This relationship between even and odd moments is also seen in the stochastic price equation, where relative growth rates increase with increasing odd moments, and decrease with increasing even moments \citep{Rice2008}.  

The relationship between moments and extinction can provide insight into the evolutionary process.  A high first moment can favor survival, but worst case extrema (``long shots'') represent the strategies that are least likely to survive.  Better strategies have a high first moment and relatively low second moment (high mean, low variance) with the worst case as the distribution with the lowest third moment (strongest right skew).  Even better strategies have a high first moment with a relatively low second moment and relatively high third moment (high mean, low variance, strong left skew).  Worst case extrema with three moments have the highest possible fourth moment (excessive kurtosis).  The relative importance of higher moments depends on the distribution, and in some cases higher moments can have a big influence on extinction.    

Strategies with a high probability of extinction are unlikely to be found in natural populations, even if their expected reproductive rate is high \citep{Orzack1980}.  New alleles will often arrive in a population as a singlet, and extinction is permanent unless the same mutation occurs more than once.  In such cases, survival is more important than the average rate of reproduction. Using moments of the offspring distribution one can find bounds on extinction using their $s$-convex extrema.  If the best case extrema for a set of moments has a high probability of extinction, then strategies with these moments will be evolutionarily unlikely, regardless of how fit they would be if they survived.

Gamblers can avoid strategies with a high risk of ruin by calculating their odds.  In natural populations, such calculations are not required to prevent the occurrence of high risk strategies.  Instead, risky strategies will be naturally unlikely, especially considering that many arrive as a single allele with one chance at survival.  Risk is not solely determined by mean growth, and strategies with a high mean can sometimes have high risk.  Unfortunately, these high risk and high reward strategies are unlikely to return anything without sufficient investment, so natural avoidance of risk can result in missed opportunity for growth.

%%%%%%%%%%%%%%%%%%%%%%%%%%%%%%%%%%%%%
\bibliography{extinction}

\begin{thebibliography}{17}
\providecommand{\natexlab}[1]{#1}
\providecommand{\url}[1]{\texttt{#1}}
\expandafter\ifx\csname urlstyle\endcsname\relax
  \providecommand{\doi}[1]{doi: #1}\else
  \providecommand{\doi}{doi: \begingroup \urlstyle{rm}\Url}\fi

\bibitem[Canjar(2007)]{Canjar2007}
M.R. Canjar.
\newblock Gambler's ruin revisited: The effects of skew and large jack- pots.
\newblock In S.N. Ethier and W.R. Eadington, editors, \emph{Optimal Play:
  Mathematical Studies of Games and Gambling}, pages 439--469. Institute for
  the Study of Gambling and Commercial Gaming, University of Nevada, Reno,
  2007.

\bibitem[Courtois et~al.(2006)Courtois, Denuit, and Bellegem]{Courtois2006}
Cindy Courtois, Michel Denuit, and Sebastien~Van Bellegem.
\newblock Discrete -convex extremal distributions: Theory and applications.
\newblock \emph{Applied Mathematics Letters}, 19\penalty0 (12):\penalty0 1367
  -- 1377, 2006.
\newblock ISSN 0893-9659.
\newblock \doi{10.1016/j.aml.2006.02.006}.
\newblock URL
  \url{http://www.sciencedirect.com/science/article/pii/S089396590600053X}.

\bibitem[Daley and Narayan(1980)]{Daley1980}
D.J. Daley and P.~Narayan.
\newblock {{S}eries expansions of probability generating functions and bounds
  for the extinction probability of a branching process}.
\newblock \emph{Journal of Applied Probability}, 17:\penalty0 939, 1980.

\bibitem[Denuit and Lefevre(1997)]{Denuit1997}
Michael Denuit and Claude Lefevre.
\newblock Some new classes of stochastic order relations among arithmetic
  random variables, with applications in actuarial sciences.
\newblock \emph{Insurance: Mathematics and Economics}, 20\penalty0
  (3):\penalty0 197 -- 213, 1997.
\newblock ISSN 0167-6687.
\newblock \doi{10.1016/S0167-6687(97)00010-3}.
\newblock URL
  \url{http://www.sciencedirect.com/science/article/pii/S0167668797000103}.

\bibitem[Denuit et~al.(1999{\natexlab{a}})Denuit, de~Vylder, and
  Lefevre]{Denuit1999b}
Michael Denuit, Etienne de~Vylder, and Claude Lefevre.
\newblock Extremal generators and extremal distributions for the continuous
  $s$-convex stochastic orderings.
\newblock \emph{Insurance: Mathematics and Economics}, 24:\penalty0 201--217,
  1999{\natexlab{a}}.

\bibitem[Denuit et~al.(1999{\natexlab{b}})Denuit, Lefevre, and
  Mesfioui]{Denuit1999}
Michel Denuit, Claude Lefevre, and Mhamed Mesfioui.
\newblock On s-convex stochastic extrema for arithmetic risks.
\newblock \emph{Insurance: Mathematics and Economics}, 25\penalty0
  (2):\penalty0 143 -- 155, 1999{\natexlab{b}}.
\newblock ISSN 0167-6687.
\newblock \doi{10.1016/S0167-6687(99)00030-X}.
\newblock URL
  \url{http://www.sciencedirect.com/science/article/pii/S016766879900030X}.

\bibitem[Ethier and Khoshnevisan(2002)]{Ethier2002}
S.~N. Ethier and Davar Khoshnevisan.
\newblock Bounds on gambler's ruin probabilities in terms of moments.
\newblock \emph{Methodology and Computing in Applied Probability}, 4:\penalty0
  55--68, 2002.
\newblock ISSN 1387-5841.
\newblock URL \url{http://dx.doi.org/10.1023/A:1015705430513}.
\newblock 10.1023/A:1015705430513.

\bibitem[H{\"u}rlimann(2005)]{Hurlimann2005}
Werner H{\"u}rlimann.
\newblock Improved analytical bounds for gambler's ruin probabilities.
\newblock \emph{Methodology and Computing in Applied Probability}, 7:\penalty0
  79--95, 2005.
\newblock ISSN 1387-5841.
\newblock URL \url{http://dx.doi.org/10.1007/s11009-005-6656-4}.
\newblock 10.1007/s11009-005-6656-4.

\bibitem[Karlin and McGregor(1957)]{Karlin1957}
Samuel Karlin and J.~L. McGregor.
\newblock The differential equations of birth-and-death-processes, and the
  stieltjes moment problem.
\newblock \emph{Transactions of the American Mathematical Society}, 85\penalty0
  (2):\penalty0 489--546, 1957.
\newblock URL \url{http://www.jstor.org/stable/1992942}.

\bibitem[Kelly(1956)]{Kelly}
J.~Kelly.
\newblock {A new interpretation of information rate}.
\newblock \emph{Bell Sys. Tech. Journal}, 35:\penalty0 917--926, 1956.

\bibitem[Kimmel and Axelrod(2002)]{Kimmel}
Marek Kimmel and David~E. Axelrod.
\newblock \emph{{Branching Processes in Biology}}.
\newblock Springer, May 2002.
\newblock ISBN 038795340X.
\newblock URL \url{http://www.worldcat.org/isbn/038795340X}.

\bibitem[Lewontin and Cohen(1969)]{Lewontin1969}
R.~C. Lewontin and D.~Cohen.
\newblock On population growth in a randomly varying environment.
\newblock \emph{Proceedings of the National Academy of Sciences}, 62\penalty0
  (4):\penalty0 1056--1060, 1969.
\newblock URL \url{http://www.pnas.org/content/62/4/1056.abstract}.

\bibitem[MacLean et~al.(2010)MacLean, Thorp, and W.T.]{MacLean2010}
L.C. MacLean, E.O. Thorp, and Ziemba W.T.
\newblock {{G}ood and bad properties of the Kelly criterion}.
\newblock In \emph{The Kelly Capital Growth Investment Criterion: Theory and
  Practice}, pages 563--574. World Scientific Publishing, Singapore, 2010.

\bibitem[Pr\'{e}kopa(1990)]{Prekopa1990}
Andr\'{a}s Pr\'{e}kopa.
\newblock The discrete moment problem and linear programming.
\newblock \emph{Discrete Applied Mathematics}, 27:\penalty0 235--254, 1990.

\bibitem[Rice(2008)]{Rice2008}
S.~H. Rice.
\newblock {{A} stochastic version of the {P}rice equation reveals the interplay
  of deterministic and stochastic processes in evolution}.
\newblock \emph{BMC Evol. Biol.}, 8:\penalty0 262, 2008.

\bibitem[Shaked and Shanthikumar(2007)]{Shaked2007}
Moshe Shaked and J.~George Shanthikumar.
\newblock \emph{Stochastic Orders}.
\newblock Springer, 2007.

\bibitem[Tuljapurkar and Orzack(1980)]{Orzack1980}
S.D. Tuljapurkar and Steven~Hecht Orzack.
\newblock Population dynamics in variable environments i. long-run growth rates
  and extinction.
\newblock \emph{Theoretical Population Biology}, 18\penalty0 (3):\penalty0 314
  -- 342, 1980.
\newblock ISSN 0040-5809.
\newblock \doi{10.1016/0040-5809(80)90057-X}.
\newblock URL
  \url{http://www.sciencedirect.com/science/article/pii/004058098090057X}.

\end{thebibliography}

%%%%%%%%%%%%%%%%%%%%%%%%%%%%%%%%%%%%%
%%% EOF
%%%%%%%%%%%%%%%%%%%%%%%%%%%%%%%%%%%%%
\end{document}